# On the Spectral Efficiency of Blind Channel Estimation and Synchronization Techniques

A. Saci, A. Al-Dweik, and A. Shami

## Abstract

In the literature, channel estimation and synchronization (CE/SY) algorithms are classified as blind, and hence spectrally efficient, if they do not require pilot symbols. However, we show in this letter that such classification is not accurate and can be misleading. Consequently, this letter presents a more reliable and accurate approach to evaluate the spectral efficiency of communications systems with various CE/SY algorithms. The proposed approach allows fair spectral efficiency comparison between various systems with blind or non-blind CE/SY algorithms. In particular, we evaluate the spectral efficiency of communications systems that incorporates blind CE/SY algorithms and compare it to other blind and pilot-based algorithms. The obtained results reveal that, on the contrary to what is widely accepted, blind CE/SY algorithms with modulation-type constrain do not necessarily improve the spectral efficiency as compared to pilot-based techniques. Consequently, such techniques are classified as conditionally blind, to distinguish them from fully blind techniques.

## I. Introduction

**B**ANDWIDTH scarcity is one of the chronic limitations that wireless networks has been suffering from since the early days of mobile data emergence. The massive and persistent increase of the traffic volume over the past few years made the network scalability processes highly challenging, frequent and costly. For example, the mobile data traffic in the second quarter of 2015 has increased by a factor of 18 as compared to the second quarter of 2010, and expected to grow by a factor of 10 between 2015 and 2020 [1]. In other words, the data growth in a single decade can be as much as 180 folds. Consequently, researchers from the academic and industrial sectors have been devoting remarkable efforts to maximize the spectral efficiency of wireless network by optimizing the spectrum utilization across all layers of the communications protocol stack.

At the physical layer (PHY), channel estimation and synchronization (CE/SY) are the key processes required to regenerate the data symbols reliably at the receiver side. Typically, CE/SY are classified based on their spectral efficiency, computational complexity, accuracy and observation window size. Based on their spectral efficiency, CE/SY techniques are categorized as blind or non-blind techniques. In general, blind techniques are considered spectrally efficient because they rely solely on the information symbols to extract the CE/SY parameters. On the

A. Saci, A. Al-Dweik and A. Shami are with the Department of Electrical and Computer Engineering, Western University, London, ON, Canada, (e-mail: {asaci, aaldweik, ashami}@uwo.ca).

contrary, non-blind, pilot-based and data-aided refer to a class of CE/SY algorithms that require reference symbols, denoted as pilots or training sequences, to extract the CE/SY parameters. Because the reference symbols are known at the receiver side, they do not carry information, and hence they degrade the system spectral efficiency. Generally speaking, pilot-based CE/SY are computationally efficient, produce reliable estimates, robust in various operating conditions, and have short observation window. Consequently, pilot-based CE/SY is adopted in most current broadband wired/wireless applications such as digital audio broadcasting (DAB) [2], digital video broadcasting-terrestrial (DVB-T) [3], Worldwide Interoperability for Microwave Access (WiMAX) technologies [4], fourth generation (4G) LTE-Advanced [5], second generation digital video transmission over cable (DVB-C2) [6], the asymmetric digital subscriber line (ADSL) [7] and home networking over power line communications (PLC) [8]. It is worth noting that all the aforementioned standards have orthogonal frequency division multiplexing (OFDM) as the modulation scheme.

Despite its many advantages, the low spectral efficiency limitation of pilot-based CE/SY techniques has motivated enormous number of researchers to develop blind techniques to overcome the spectral efficiency problem. For example, various blind carrier frequency offset (CFO) estimation, blind symbol timing offset (STO) estimation and blind channel estimation algorithms are reported in [9]-[13], [14], and [15]-[29], respectively. However, many of the blind CE/SY techniques reported in the literature have a strict constraint regarding the modulation type [9]-[29]. In particular, constant modulus (CM) constellations are required for such algorithms to function properly.

Typically, the spectral efficiency for OFDM systems is computed as $(N - N_P)/N$, where $N$ is the total number of subcarriers and $N_P$ denotes the number of subcarriers allocated for pilot symbols [30]. For example, the spectral efficiency of LTE-A is about %95, and it is about %91.6 for DVB-T, and hence, significant spectral gain can be achieved by adopting blind techniques. However, spectral efficiency comparisons in most of the work reported in the literature is based on the assumption that all subcarriers in the concerned systems are using the same modulation type and order. Consequently, such comparison is valid only when the modulation type and order similarity condition is satisfied. However, pilot-based CE/SY algorithms generally have no constraints on the modulation type, and hence, the spectral efficiency comparison becomes unfair.

In this letter, we present a more informative and fair approach for spectral efficiency comparisons among various blind and pilot-based techniques, where each system has its own constraints. The new approach is based on computing the spectral efficiency for all considered systems under unified constraints, which are typically imposed in practical systems to meet certain predefined QoS requirements. On the contrary to what is widely believed, the obtained results show that several blind techniques might actually be less specially efficient than pilot-based CE/SY techniques for a broad range of signal-to-noise ratios (SNRs).

The rest of the letter is organized as follows. Section II presents the OFDM system and channel models as well as the proposed spectral efficiency evaluation approach. Section III presents the numerical examples, and finally

Section IV concludes the letter.

## II. System and Channel Models

### A. OFDM System and Channel models

In OFDM systems, a sequence of $N$ complex data symbols is used to modulate $N$ orthogonal subcarriers during the $\ell$th OFDM block $\mathbf{d}(\ell) = [d_0(\ell), d_1(\ell), ..., d_{N-1}(\ell)]^T$. However, $N_P$ data symbols, denoted as pilots, do not actually carry information because they are known at the receiver side. The data symbols $d_k$, including the pilots, are usually drawn uniformly from a quadrature amplitude modulation (QAM), phase shift keying (PSK) or amplitude shift keying (ASK) constellation. The sequence of data and pilot symbols is modulated using an $N$-point inverse discrete Fourier transform (IDFT) process that produces the sequence $\mathbf{x}(\ell) = [x_0(\ell), x_1(\ell), \cdots, x_{N-1}(\ell)]^T$. Thus

$$\mathbf{x}(\ell) = \mathbf{F}^H \mathbf{d}(\ell) \tag{1}$$

where $\mathbf{F}$ is the normalized $N \times N$ DFT matrix. Then, the cyclic prefix (CP) is created by copying the last $N_g$ samples of the IDFT output and appending them at the beginning of the symbol to be transmitted. Therefore, the transmitted OFDM block consists of $N_t = N + N_g$ samples. The useful part of the OFDM symbol does not include the $N_g$ prefix samples and has a duration of $T_u$ seconds.

At the receiver front-end, the received signal is applied to a matched filter and is then sampled at a periods $T_s = T_u/N$. Assuming that the channel is fixed within one OFDM symbol, dropping the CP samples, and applying the DFT to the received sequence gives,

$$\mathbf{y}(\ell) = \mathbf{H}(\ell)\mathbf{d}(\ell) + \mathbf{z}(\ell) \tag{2}$$

where $\mathbf{H}(\ell)$ denotes the channel frequency response during the $\ell$th OFDM block

$$\mathbf{H}(\ell) = diag\left([H_0(\ell), H_1(\ell), \ldots, H_{N-1}(\ell)]^T\right)$$

and $\mathbf{z}(\ell) = [z_0(\ell), z_1(\ell), \cdots, z_{N-1}(\ell)]^T$ denotes the additive system noise, which is modeled as a white Gaussian process with zero mean and variance $\sigma_z^2$.

To maximize the efficiency of OFDM-based communication systems, the modulation types/orders of the information symbols in $\mathbf{d}(\ell)$ are chosen based on the channel matrix $\mathbf{H}(\ell)$ [31], which is assumed to be known at the transmitter side via a feedback channel, and the instantaneous SNR,

$$\gamma_k = \frac{|H_k|}{\sigma_z^2} |d_k|^2. \tag{3}$$

However, to minimize the signaling over the feedback channel, and to exploit the time/frequency correlation of the channel, the channel information is grouped into blocks, each of which has $N_F$ subcarriers in frequency domain and $N_T$ subcarriers in time domain, which forms one resource block of size $N_B = N_F \times N_T$. Therefore, all subcarriers within a particular block are assigned to the same modulation type/order. In LTE, the resource block for FDD 1.4 MHz with normal CP has $N_F = 12$ and $N_T = 7$, and hence $N_B = 84$. More generally, each subcarrier can be modulated using different modulation type and order.

TABLE I
$M_{n,m}$ EXAMPLE FOR N=1,2,3 AND DIFFERENT VALUES OF $M$.

| $n \setminus m$ | 1 | 2 | 3 | 4 | 5 | 6 | 7 |
|---|---|---|---|---|---|---|---|
| 1 (ASK) | 1 | 2 | 4 | 8 | — | — | — |
| 2 (PSK) | 1 | 2 | 4 | 8 | 16 | — | — |
| 3 (QAM) | 1 | — | 4 | — | 16 | — | 64 |

## B. Spectral Efficiency of OFDM Systems

Generally speaking, the spectral efficiency $\zeta$ of OFDM based systems is usually computed as the ratio of the number of data-bearing subcarriers to the total number of subcarriers, and thus

$$\zeta = 1 - \frac{N_P}{N}. \tag{4}$$

However, such definition is valid only when all subcarriers in both systems are modulated using the same modulation type and order. In practice, different subcarriers can be modulated using different modulation schemes and orders. Therefore, the relative spectral efficiency between two OFDM-based systems should be computed as the ratio between the total number of information bits of the first system to those in the second system over one information (resource) block [5]. Therefore, we can define the relative spectral efficiency as,

$$\eta_R = \frac{\eta^{(1)}}{\eta^{(2)}} = \frac{\sum_{\ell=0}^{N_T-1} \sum_{k=0}^{N_F-1} \log_2 [M_\mathbf{v}], \mathbf{v} \in \mathcal{M}^{(1)}}{\sum_{\ell=0}^{N_T-1} \sum_{k=0}^{N_F-1} \log_2 [M_\mathbf{v}], \mathbf{v} \in \mathcal{M}^{(2)}} \tag{5}$$

where $M_\mathbf{v}$ is the modulation order for given configuration $\mathbf{v} = [n, m, k, \ell]$, where $n$ and $m$ denote the modulation type and order sequence number, respectively, $k$ and $\ell$ denote the subcarrier index in frequency and time, respectively. The set $\mathcal{M}^{(.)}$ is the set of all possible values of $n$ and $m$ for a particular system. For example, assume that a particular system supports three different modulation schemes with different modulation orders as depicted in Table I, and the time-frequency grid has $N_F = 4$ and $N_T = 4$. Consequently, the modulation map $\mathcal{M}$ will have the structure given in Table II. As it can be noted from the table, the four subcarriers in the corners of the table carry no information because $m = 1$, the remaining subcarriers in row-0 and row-3 are limited to MPSK, but with any order. The subcarriers in the first row can have any combination of $\{n, m\}$, and the subcarriers in the second row are limited to ASK modulation with any $m$ value. .

.

In practical systems, the map $\mathcal{M}$ is specified at the initial stages of the system design, and then, the values of $n$ and $m$ are dynamically selected based on the system QoS requirements, the system resources, and the channel state information (CSI) [32]. Without loss of generality, consider the case where the values of $m$ and $n$ can be selected dynamically with the aim of maximizing the spectral efficiency of a particular system under bit error rate (BER), and modulation type/order constraints. Therefore, the problem can be formulated as

$$\max_{m,n} \sum_{\ell=0}^{N_T-1} \sum_{k=0}^{N_F-1} \log_2 [M_\mathbf{v}] \tag{6}$$

subject to:

$$[n,m] \in \mathcal{M} \tag{7a}$$

$$\bar{P} \leq P_T \tag{7b}$$

where (7a) is used to guarantee that the system uses only the allowed modulation types and orders, and (7b) is used to guarantee that the average BER $\bar{P}$ is less than a prescribed threshold $P_T$,

$$\bar{P} = \frac{\sum_{\ell=0}^{N_T-1} \sum_{k=0}^{N_F-1} \log_2 (M_\mathbf{v}) P_\mathbf{v}(\gamma_k)}{\sum_{\ell=0}^{N_T-1} \sum_{k=0}^{N_F-1} \log_2 (M_\mathbf{v})} \leq P_T \tag{8}$$

where $P_\mathbf{v}(\gamma_k)$ is the instantaneous BER given $\mathbf{v}, \gamma_k$. In typical bit loading problems, $\bar{P}$ is computed with the assumption of perfect CSI knowledge at the transmitter. In spectral efficiency analysis, the accuracy of the algorithm, SNR and spectral efficiency are correlated. For example, two blind CE algorithms with different accuracy would actually have different spectral efficiency.

TABLE II
Time-frequency modulation map example.

| $\ell \backslash k$ | 0 | 1 | 2 | 3 |
|---|---|---|---|---|
| 0 | $2,1$ | $2,m$ | $2,m$ | $2,1$ |
| 1 | $n,m$ | $n,m$ | $n,m$ | $n,m$ |
| 2 | $1,m$ | $1,m$ | $1,m$ | $1,m$ |
| 3 | $2,1$ | $2,m$ | $2,m$ | $2,1$ |

## III. Numerical Results

In this section, simulation results are presented to evaluate the average throughput per subcarrier and the relative spectral efficiency. The channel is assumed to be frequency-selective quasi-static with Rayleigh fading, where the channel remains fixed within one OFDM symbol, but changes randomly over consecutive symbols. The channel model considered in this work is the typical urban (Tux) multipath fading channel model [33] that consists of 9 taps with normalized delays of $[0, 1, ..., 8]$ and average normalized taps' gains of $[2.69, 1.74, 2.89, 1.17, 0.23, 0.58, 0.36, 0.26, 0.08]/10$.

The spectral efficiency for four different systems is considered, which are the fully blind (FB), CM, LTE and the modified LTE (M-LTE) [35]. The FB system is similar to LTE except that no pilots are used. The CM has no pilots, but the modulation is limited to MPSK. The M-LTE is similar LTE except that pilot symbols are replaced by unipolar $M$-ary amplitude shift keying (MASK) and one of the subcarriers adjacent to pilot should have CM as well. Moreover, the modulation order $M_v$ can be set to one to satisfy the BER requirements. In all the considered systems, the appropriate modulation order is selected such that the average BER is less than $P_T$. The modulation orders for all subcarriers are computed using the Incremental Allocation Algorithm proposed in [34]. The spectral efficiency of the FB system is considered as $\eta^{(2)}$ when $\eta_R$ is computed, because FB has the maximum spectral efficiency.

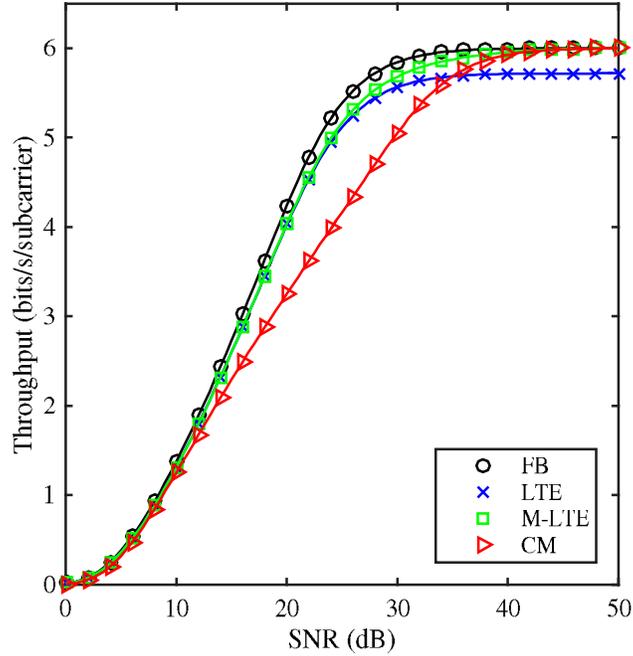

Fig. 1. Throughput per subcarrier at $P_T = 10^{-3}$.

Fig. 1 presents the average throughput per subcarrier, for the FB, CM, LTE, and M-LTE [5] systems. As it can be noted from the figure, the FB system outperforms all other systems since it does not require pilots, and it has no modulation-type constraint. Unlike what is usually assumed, the LTE outperforms CM systems for a wide rang of SNRs. Therefore, sacrificing a few subcarriers as pilots and selecting the modulation type for other subcarriers freely results in higher throughput as compared to the case where all subcarriers carry information, but have the CM constraint. The M-LTE throughput is equivalent to LTE at low SNRs, but it shows higher throughput at high SNRs.

The relative spectral efficiency $\eta_R$ of the CM, LTE and M-LTE systems is presented in Fig. 2 for BER thresholds $P_T = 10^{-2}, 10^{-3}$ and $10^{-4}$. As it can be noted from the figure, the LTE system has a constant spectral efficiency of about $95\%$, where the $5\%$ loss is caused by the pilots. Surprisingly, the figure shows that LTE outperforms the blind M-LTE and CM system for low to medium SNRs. The figure also shows that the spectral efficiency is depends on the SNR, $P_T$ because both parameters affect the selection of the modulation orders for the different systems. For example, in the range of low SNRs, we note that $\eta_R$ for the CM system is increasing as a function of the SNR, which is due to the fact that the majority of the FB system subcarriers at this range of SNRs are modulated mostly using $M = 1, 2$ and $4$, which is similar to the CM case. In the mid-rage SNRs, more subcarriers in the FB system will start to use 16-QAM, while the CM is mostly limited to $M \leq 8$, and hence $\eta_R$ decreases. Finally, at high SNRs, the FB will be mostly using 64-QAM, which is the maximum allowed modulation order, and hence $\eta^{(1)}$ of the CM will eventually approach $\eta^{(2)}$ of the FB system. Similar to the CM case, the LTE outperforms the M-LTE at low SNRs, however, the difference is negligible. At high SNRs, the M-LTE outperforms the LTE noticeably.

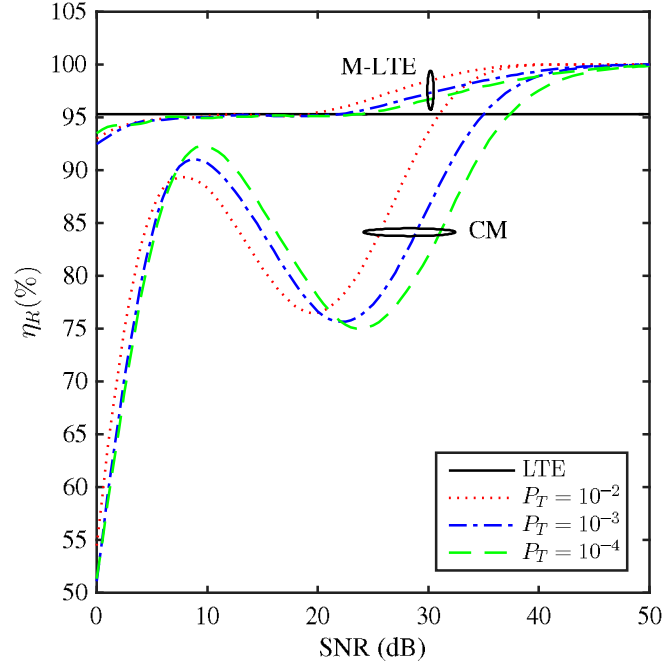

Fig. 2. Relative spectral efficiency for an OFDM system with different modulation constraints.

## IV. CONCLUSIONS

In this letter, the concept of spectral efficiency of blind CE/SY techniques is revisited, where we proposed a new fair and reliable approach to compare the spectral efficiency of various blind and non-blind communications systems. The new approach considers the fact that different subcarriers in OFDM systems may be modulated using different modulation types and orders to satisfy QoS requirements. Moreover, the proposed approach considers the modulation type constraint on the overall system special efficiency. The obtained results showed that the modulation type constraint has a significant impact on the system spectral efficiency, which can make the spectral efficiency of pilot-based systems higher than that of the blind with modulation type constraint.